\documentclass[%
 reprint,
superscriptaddress,
 amsmath,amssymb,
 aps,
prl,
]{revtex4-2}

\usepackage{graphicx}
\usepackage{dcolumn}
\usepackage{bm}
\graphicspath{ {images/} }
\usepackage{textgreek}
\usepackage[T1]{fontenc}
\usepackage[utf8x]{inputenc}

\usepackage[colorinlistoftodos]{todonotes}

\usepackage[colorinlistoftodos]{todonotes}

\usepackage[colorinlistoftodos]{todonotes}

\begin{document}

    \title{ Advanced magnon-optic effects with spin-wave leaky modes }

    \author{Krzysztof Sobucki}
    \email{krzsob@st.amu.edu.pl}
    \affiliation{Institute of Spintronics and Quantum Information, Faculty of Physics, Adam Mickiewicz University, Uniwersytetu Poznańskiego 2, 61-614 Poznań, Poland}
    
    \author{Wojciech Śmigaj}
    \affiliation{Met Office, FitzRoy Rd, Exeter, EX1 3PB, UK}
    
    \author{Piotr Graczyk}
    \affiliation{Institute of Molecular Physics, Polish Academy of Sciences, Mariana Smoluchowskiego 17, 60-179 Poznań, Poland}
    
    \author{Maciej Krawczyk}
    \affiliation{Institute of Spintronics and Quantum Information, Faculty of Physics, Adam Mickiewicz University, Uniwersytetu Poznańskiego 2, 61-614 Poznań, Poland}
    
    \author{Paweł Gruszecki}
    \email{gruszecki@amu.edu.pl}
    \affiliation{Institute of Spintronics and Quantum Information, Faculty of Physics, Adam Mickiewicz University, Uniwersytetu Poznańskiego 2, 61-614 Poznań, Poland}

    \begin{abstract}
    
    We numerically demonstrate the excitation of leaky spin waves (SWs) guided along a ferromagnetic stripe by an obliquely incident SW beam on the thin film edge placed below the stripe. During propagation, leaky waves emit energy back to the layer in the form of plane waves and several laterally shifted parallel SW beams. This resonance excitation, combined with interference effects of the reflected and re-emitted waves, results in the magnonic Wood's anomaly and significant increase of the Goos-Hänchen shift magnitude. Hence, we provide a unique platform to control SW reflection and to transfer SWs from a 2D platform into the 1D guiding mode that can be used to form a transdimensional magnonic router.
    
    \end{abstract}
    

    \maketitle

    In wave physics, bound and extended modes can be recognised due to their amplitude spatial distribution. 
    The most common are the extended states, which freely propagate in the system. Examples of the second type  include bound states in the continuum (BICs) and leaky modes (LMs)\cite{Hsu2016}.
    BIC is a state, which exists in the continuous part of the spectra but is perfectly localised, it was predicted  by von Neuman and Wigner for electron waves~\cite{Neuman1929} and later experimentally observed  for photons \cite{capasso1992observation, hsu2013observation,Bykov2020} and phonons \cite{stegeman1976normal, aleksandrov1992mandelstamm}.
    LMs are another type of modes, which are localised but can store energy only for a limited time due to their coupling with extended states. Therefore, LMs can be excited by the propagating modes, and also leaks energy into them. Hence, the LM wavevector is a complex number, which the imaginary part specifies the rate of energy leakage~\cite{Moiseyev, Kukulin, tamir1971lateral}. 
    The LMs facilitate the occurrence of the Wood's anomaly that was first reported for light reflected from the grating~\cite{wood1902,hessel1965Woods, tamir1971lateral}.

    An intriguing wave type is a spin wave (SW) that is, a collective precessional disturbance of magnetisation in magnetic materials, which is believed to be a promising candidate for information carriers in beyond-CMOS devices
    \cite{BeyondCMOS, Manipatruni2018, csaba2017, chumak2021roadmap}.
    SW optics is more complex than its electromagnetic counterpart and rich in optical phenomena\cite{hoefer,gruszecki2016microwave,heussner2018frequency, stigloher2016snell, hioki2020snell, yu2016magnetic, Golembiewski2020,golebiewski2022LUT, Mieszczak2020}. 
    Many effects from photonics have already been transferred to magnonics, for instance, negative refraction\cite{kim2008}, anomalous refraction\cite{Mieszczak2020}, mirage effect\cite{gruszecki2018}, and the Goos-Hänchen (GH) effect\cite{goos1947}, i.e., the lateral shift of the waves's reflection point at the interface \cite{gruszecki2014goos,gruszecki2015influence,Stigloher_2018,laliena2021magnonic, zhen2020giant}. Whilst GH effect was predicted theoretically, it has not yet been experimentally observed for SW beams. Also the BICs  \cite{Yang2020}, LMs, Wood anomalies, and the resonance effects widely explored in photonic \cite{yu2022active} remain poorly present in magnonics.

\begin{figure}
        \centering
        \includegraphics[width=8.6cm]{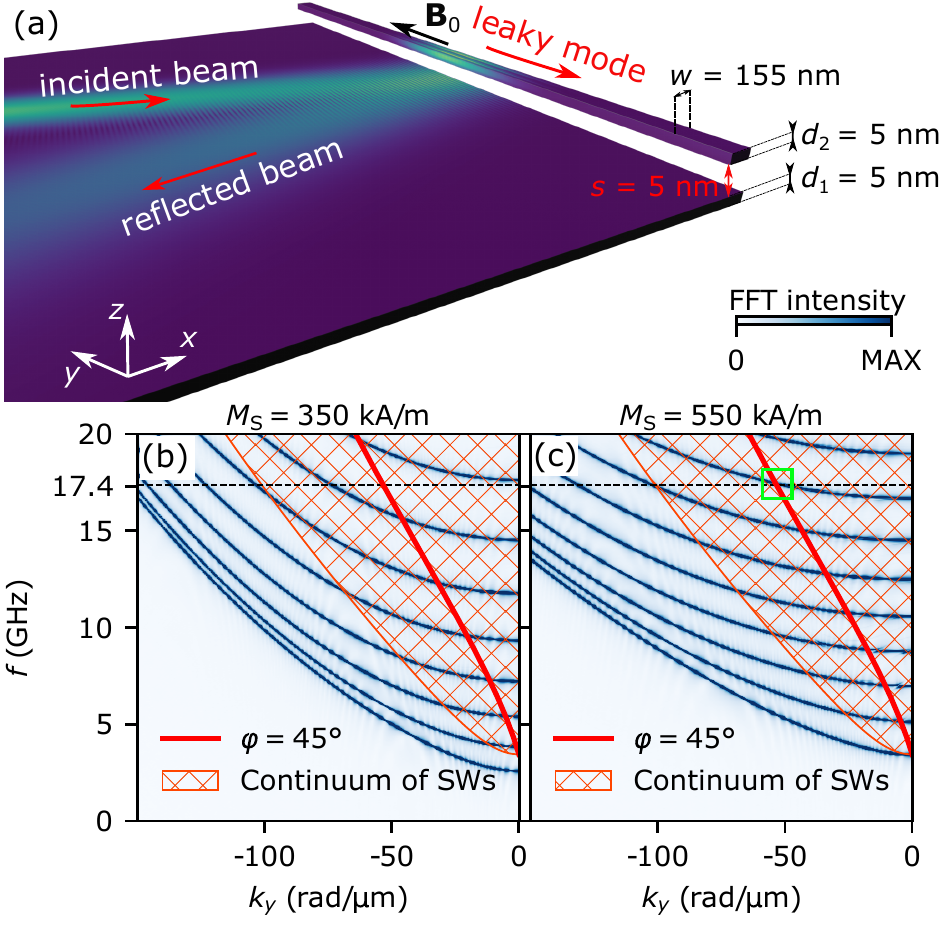}
        \caption{
        (a) Geometry of the system. Colour represents the SW amplitude and indicates the incident and reflected SW beams in the film, as well as the LM in the stripe.
        (b) and (c) The dispersion relation of SWs. The blue colourmaps in the background represent dispersion relations of the resonant-stripe element of the SWs propagating along the stripe with (b) $M_\mathrm{S}=350$~kA/m and (c) $M_\mathrm{S}=550$~kA/m.
        The hatched region displays the continuum of SWs in the CoFeB layer. The thick red line represents the analytical dispersion relation in dependence on $k_y$ for SWs propagating in the film at an angle $\varphi=45^{\circ}$ to $\mathbf{B}_0$. The horizontal black-dashed line indicates the frequency $f_0=17.4$~GHz used in steady-state simulations. 
        The green square in (c) shows the crossing at $f_0=17.4$ of the beam and bilayer dispersions.
        } 
        \label{fig:geo} 
    \end{figure}

    The use of nanoresonators in the form of ferromagnetic stripes placed over a thin film to modulate SWs has recently attracted attention~\cite{kruglyak2017graded,au2012phaseshifter,au2012transducer, al2008evidence, zhang2019resonator, fripp2021darkModes, sobucki2021, sobucki2022IEEE, smigaj2021modal, qin2021nanoscale}.
    In this letter, we numerically study the oblique reflection of a SW beam from the edge of the ferromagnetic film ending with the resonant-stripe element~\cite{sobucki2021} [see Fig.~\ref{fig:geo}(a)].  
    We find that an SW beam can excite an LM when the resonant conditions are met. The LM emits SWs back to the film while propagating along the stripe. As a result, we observe an anomaly in the amplitude of reflected waves, which we interpret as a magnonic counterpart of the Wood's anomaly. Moreover, we detect multiple reflected beams with tunable positions. Thus, the excitation of LMs allows tuning the GH shift by several wavelengths. Our results open a route for the exploitation of the demonstrated effects in various magnonic applications, including designing resonant SW metasurfaces in planar structures suitable for integration with magnonic devices, converting extended SWs into a guided stripe wave, and proposing a method for exploiting 3rd dimension in integrated magnonic systems.

    We consider a  half-infinite CoFeB layer of thickness $d_1=5$~nm with the saturation magnetization 1200~kA/m and the exchange constant 15~pJ/m.
    Above the film, a ferromagnetic stripe of width $w=155$~nm and thickness $d_2=5$~nm aligned with the layer's edge is placed. We assume that the exchange constant in the stripe equals to $3.7$~pJ/m and the value of $M_\mathrm{S}$ varies.
    Both elements are separated by $s=5$~nm dielectric nonmagnetic layer, see Fig.~\ref{fig:geo}(a). Let us refer to the stripe and the layer directly below it as a bilayer.
    The system is uniformly magnetised by the external magnetic field $B_0 = \mu_0 H_0=0.1$~T directed along the stripe (the $y$-axis). We set the damping parameter $\alpha=0.0004$ and the gyromagnetic ratio $\gamma=-176$~radGHz/T. 
    Under the normal SW incidence, this geometry is a magnonic realisation of  the Gires-Tournois interferometer offering multiple Fabry-Pérot resonances~\cite{sobucki2021}. We analyse the oblique incidence of a $775$~nm wide SW beam at a frequency $f_0=17.4$~GHz (wavelength $\lambda=103$~nm)  under an angle of  45$^\circ$\footnote{the angle of the phase velocity with respect to the $x$-axis}. We employ MuMax3\cite{vansteenkiste2014design}
    to perform micromagnetic simulation of magnetization $\mathbf{m}(\mathbf{r},t))$ dynamics in the system (for more details see supplementary material).

    The dispersion relations of SWs propagating along the stripe placed above the layer for the two selected values of stripe's $M_\mathrm{S}$, i.e., $350$~kA/m and $550$~kA/m, are shown by the blue colourmaps in Fig.~\ref{fig:geo}(b) and (c), respectively. 
    The hatched area is the continuum spectrum of propagating SWs calculated analytically \cite{kalinikos1986theory} in the CoFeB film, and the red line is the dispersion of the SW beam (see supplementary material). 
    For $M_\mathrm{S}=550$~kA/m, the analytical dispersion crosses the bilayer dispersion at the frequency  of the SW beam $f_0=17.4$~GHz, cf. green square in Fig.~\ref{fig:geo}(c). Here, the wavevector component $k_y$ of the incident wave matches the wavenumber of the bilayer mode, so we expect the incident SW beam to excite the mode in the bilayer. For $M_\mathrm{S}=350$~kA/m, there is no phase matching at $f_0=17.4$~GHz; thus, the coupling between the incident SW and the bilayer mode is suppressed.

    \begin{figure}
        \centering
        \includegraphics[width=8.0cm]{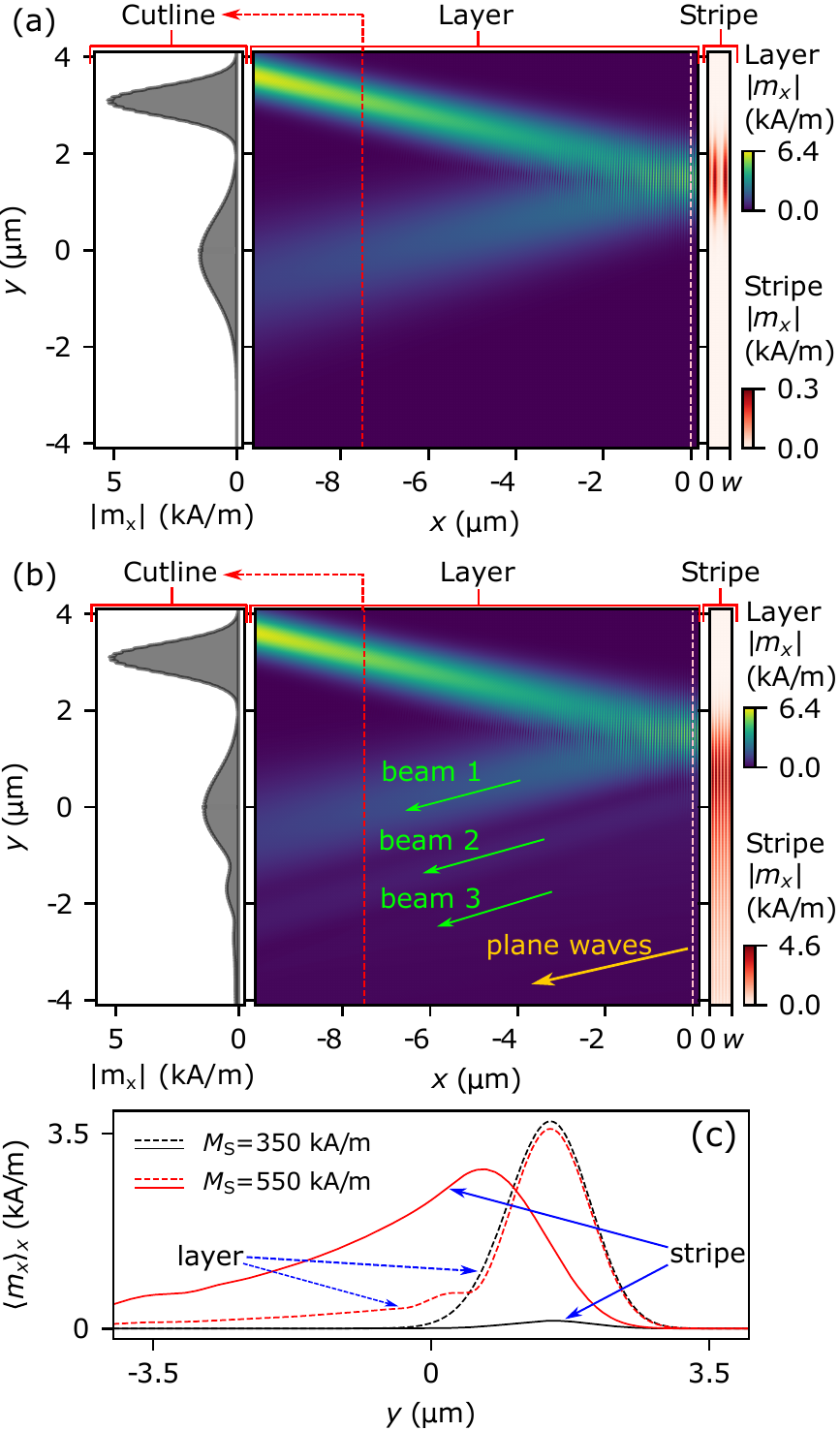}
        \caption{
        Colourmaps of the SW amplitude distribution of magnetization's in-plane dynamic component $|m_x|$. Middle panels show $|m_x|$ in the CoFeB layer. The colourmaps in the bars on the right present $|m_x|$ in the stripe. The panels on the left show $|m_x|$  in the layer in a cutline through in the layer at $x=-7.5$ \textmu m indicated with red-dashed line in the central panel. 
        Results for the system with stripe of (a)  $M_{\mathrm{S}}=350$~kA/m and (b) $M_{\mathrm{S}}=550$~kA/m. (c) Averaged $\langle |m_x| \rangle_x(=\frac{1}{w}\int_{0}^{w}|m_x| dx)$ in dependence on $y$ in the stripe (solid lines) and in the layer directly under the stripe (dashed lines). The black and red lines represent results for the system with the stripe $M_{\mathrm{S}}=350$~kA/m and $M_{\mathrm{S}}=550$~kA/m, respecitvely.
        } \label{fig:maps} 
    \end{figure}

    To verify our predictions, we examine the reflection of the SW beam from the bilayer edge for the two considered values of $M_\mathrm{S}$ in the stripe. In the simulations, we use a continuous excitation of the SW beam and analyse the steady-state SW distribution in the system (see supplementary material). Figs.~\ref{fig:maps} (a) and (b) present the steady state $|m_x|$ amplitude distributions for $M_\mathrm{S}$ in the stripes 350~kA/m and 550~kA/m, respectively. 
    
    In the case of the stripe $M_\mathrm{S}=350$~kA/m [Fig.~\ref{fig:maps} (a)] the SWs are excited, but oscillations are present only in the region above the incident spot, and its amplitude is one order of magnitude smaller than that in the layer. In the far field,
    we see only a single reflected SW beam, cf. the left panel in Fig.~\ref{fig:maps}(a). 
    Both the incident and reflected beams have Gaussian envelopes. 
    
    We observe a different behaviour in the case of $M_\mathrm{S}=550$~kA/m. First, the amplitude of stripe's SWs is comparable with the SW beam in the layer.
    Moreover, we observe the propagation of SWs along the stripe. 
    The mode in bilayer emits SWs back to the layer during its propagation, what is a clear indication of its LM nature.
    Furthermore, we observe the formation of new SW beams in the layer that are parallel to the primary beam, see Fig.~\ref{fig:maps}(b). 
    Two new beams are clearly visible in the left panel presenting an SW intensity cross-section taken at $x=-7.5$ \textmu m. Note that, there are also plane waves propagating outwards the interface.

    Let us examine the change of the SW amplitude as it propagates along the bilayer. In Fig.~\ref{fig:maps}(c)
    we see a significant difference between the SW modes in the two considered cases.
    As expected, for the stripe $M_{\mathrm{S}}=350$~kA/m, the SW amplitude in the stripe is negligible.
    However, when the phase-matching condition is fulfilled for $M_\mathrm{S}=550$~kA/m, we see the efficient excitation of the LM that propagates along the $-y$-direction 
    (see the solid red line in Fig.~\ref{fig:maps}(c)). 
    Note that, the amplitude $\langle |m_x| \rangle_x$ in the layer below the resonator along the y-axis can be decomposed on several superimposed Gaussian functions (discussed in details in the following paragraph), namely, a  dominant one associated with the primary reflected SW beam and several additional ones with smaller amplitudes (see the dashed red line). This opens the channels for the energy leakage from the LM to the film.
    
    For a deeper insight into how LM emission occurs over time,
    let us perform simulations with $45^{\circ}$ incidence of a SW packet with the full width at half maximum (FWHM) equal to  $0.5$~ns instead of the steady-state simulations (see supplementary material). 
    In Fig.~\ref{fig:packet}, we present two snapshots from the simulations with reflected wavepacket in the system with the stripe $M_{\mathrm{S}}=~550$~kA/m (the video in supplementary material, Movie S5.).
    These simulations confirm that the leaky mode excited by the incident SWs propagates along the bilayer and reemits SWs in the form of plane waves without constant supply of SWs from the incident SW beam. In addition, it is shown in Movie S5. that the amplitude of the excited mode in the stripe bounces obliquely between the edges of the stripe (it is visible for lower amplitudes of the SWs). We stipulate that this bouncing in the stripe is the source of the spatial shift of the third and higher reflected beams. The bouncing mode reemits its energy at a higher rate when it reflects from the left edge of the resonator, giving rise to new reflected beams in the system.

    To understand the formation of the multiple beams in the reflection at resonance [Fig.~\ref{fig:maps} (b)] we perform the analysis proposed by Tamir and Bertoni\cite{tamir1971lateral} for electromagnetic waves where creation of an additional reflected beam is described. 
    The reflectance coefficient of the incident SW beam at an interface with LM can be described as $\rho(k_y)=e^{i\Delta}(k_y-k_\mathrm{p}^{\ast})/(k_y-k_\mathrm{p})$ where $\Delta$ is the phase shift between the incident and reflected beams, $k_y$ is the tangential component of the incident wavevector, and $k_\mathrm{p}=\kappa + i \nu$ is the LM wavenumber. As the Tamir-Bertoni model shows, when the stripe mode has a bound (non-flux with $\nu = 0$) character, LM is not excited by the incident beam; thus, it has no effect on the reflected beam. Excitation of the stripe mode occurs when it takes on a leaky character and is more effective for the larger imaginary part of $\nu$. Concurrently, an increase in $\nu$ accelerates the transfer of energy back to the layer; hence, at a specific $\nu$ value,  the secondary beam overshadows the main reflected beam.
    This qualitatively reflects our simulation results  (see supplementary material).    
    However, our simulations show the presence of several reflected beams instead of two, as in the Tamir-Bertoni model. This discrepancy may be due to some differences between Tamir-Bertoni model and our system such as neglecting the higher-order poles of the reflection coefficient in the Tamir-Bertoni model, the finite stripe width and bouncing of the SW amplitude between the edges of the stripe described in the previous paragraph.

    \begin{figure}[t]
        \centering
        \includegraphics[width=8.6cm]{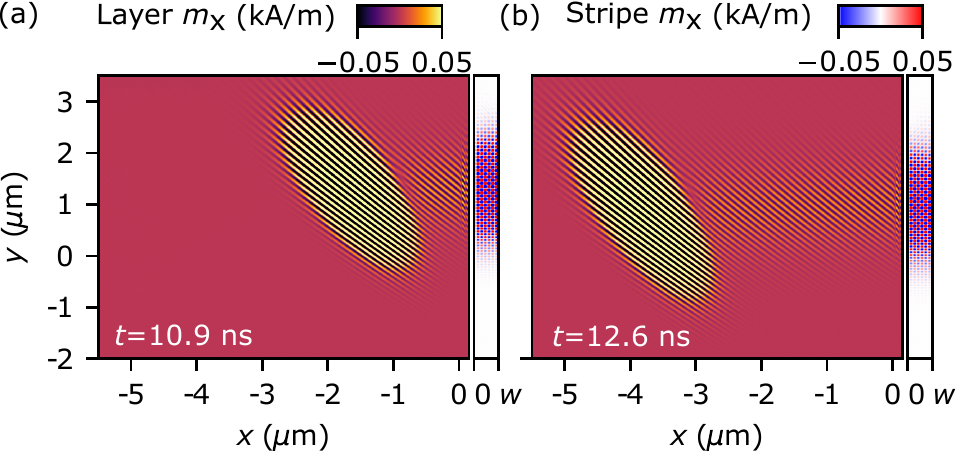}
        \caption{
        The reflection of SW wavepacket on the bilayer resonant-stripe element ($M_\mathrm{S}$=550~kA/m) for (a) $t=10.9$ ns and (b) $t=12.6$ ns. Note that the amplitudes of SWs are amplified to better visualize SWs reemited by leaky mode, see the colourbars.
        } \label{fig:packet} 
    \end{figure}

      \begin{figure}
        \centering
        \includegraphics[width=8.1cm]{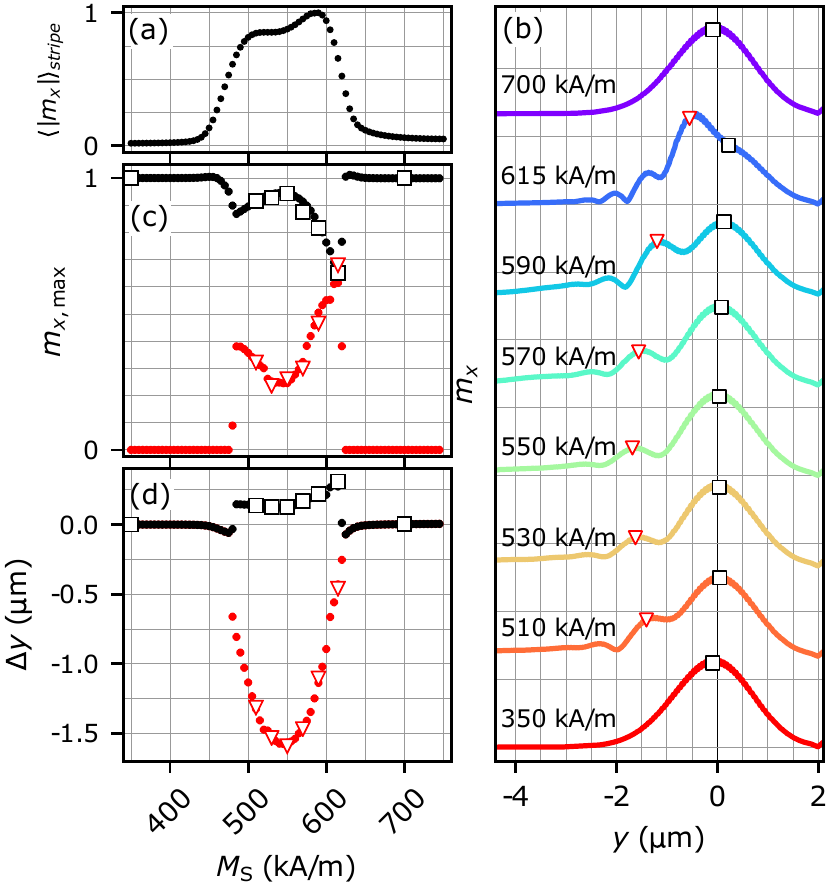}
        \caption{
        (a) Averaged $|m_x|_{stripe}$ over stripe's volume as a function of stripe's $M_{\mathrm{S}}$. (b) Comparison of several reflected beam cutlines at $x=-7.5$ \textmu m for different values of stripe's $M_{\mathrm{S}}$ where maximal amplitudes of the primary and secondary beams are marked with black squares and red triangles, respectively. (c) The maximal amplitudes of the primary and secondary beams as a function of $M_{\mathrm{S}}$ in arbitrary units. 
        (d) The spatial GH shift of beams' maxima positions with respect to the system with the stripe $M_{\mathrm{S}}=350$~kA/m as a function of $M_{\mathrm{S}}$.
        } \label{fig:shift} 
    \end{figure}

    Let us examine how the reflection is affected by the change of the stripe's $M_{\mathrm{S}}$ while going through resonance.
    Fig.~\ref{fig:shift}(a) illustrates the dependence of the average SW amplitude $\langle |m_x| \rangle_\mathrm{stripe}$ in the stripe's volume 
    on $M_{\mathrm{S}}$ in the range 350-750~kA/m.
    It shows an increase of the amplitude for $M_{\mathrm{S}} \in (450-650)$~kA/m with a maximum reaching at 590~kA/m.
    Therefore, band crossing occurs near $17.4$ GHz for a wide range of stripe's $M_{\mathrm{S}}$. The origin of this broad-band resonance effect together with dispersion relations for various $M_{\mathrm{S}}$ can be found in supplementary material.

    In Fig.~\ref{fig:shift} (b) we superimpose several cutlines through layer's $|m_x|$ distributions in the far field for different stripe's $M_{\mathrm{S}}$. We quantify the positions and amplitudes of reflected beams by fitting Gaussian curves to the cutlines in the far field, more details in supplementary material. We mark the positions of the maximum of the primary reflected beams with a black square and the secondary beams with a red triangle. The position $y=0$ represents the position of the reflected beam for the stripe $M_{\mathrm{S}}=350$~kA/m at $x=-7.5$~\textmu m.
    It is evident that depending on the $M_\mathrm{S}$ value, the positions and amplitudes of the beams change. 
    As we approach resonance, the primary beam's amplitude decreases when LM's excitation is observed, see Fig.~\ref{fig:shift}(c), decreasing by almost 40\% at $M_{\mathrm{S}}\approx 615$~kA/m. 
    We interpret such a decrease 
    as a magnonic analogue to the Wood's anomaly\cite{wood1902} since the amplitude of the reflected beam decreases due to the excitation of stripe's localised mode. 
    As the primary beam amplitude decrease, the amplitude of the secondary beam increases, and for $M_\mathrm{S} \approx 615$~kA/m it is even higher than the amplitude of the primary beam. These facts adequately reflect the Tamir-Bertoni analytical model and indicate an increase of the imaginary part of the LM wavenumber, meaning that more energy is leaked by the LM, as well as that the energy of the incident beam is more efficiently directed to the secondary reflected beams.

    As shown in Fig.~\ref{fig:shift}(d), the reflected beams are shifted with respect to the reference nonresonant scenario, i.e., $M_{\mathrm{S}}=350$~kA/m. It implies the possibility of manipulating the value of the GH shift, which for $M_{\mathrm{S}}=350$~kA/m takes the value of $+12$~nm, namely, around 10\% of the incident SW wavelength. It is a typical value of GH shifts for SWs, which are usually smaller than the SW wavelength\cite{gruszecki2014goos,gruszecki2015influence,laliena2021magnonic}.
    The positions of the primary and secondary reflected beams change when LMs are excited, cf. Figs.~\ref{fig:shift}(a) and (d).
    The change of the GH shift for the primary and secondary beams moves towards
    positive and negative $y$-coordinates, respectively.
    The displacement of the primary beam spans from $-71$~nm to $300$~nm, reaching a maximum at  $M_\mathrm{S}=615$ kA/m. 
    Therefore, with the addition of the stripe above the edge and resonance effect, we can significantly enhance and manipulate the value of the GH shift, reaching the value of up to almost three wavelengths. This is already a measurable value that is essential for the experimental verification of the GH shift for SW beams. Moreover, the shift of the secondary beam is even larger and reaches up to $-1600$~nm. Noteworthy, for $M_\mathrm{S} \approx 615$~kA/m (the case where the amplitude of the secondary beam is greater than the amplitude of the primary beam), the lateral shift value is $-456$ nm.

    In conclusion, we have shown a new way to control SWs propagation in a thin ferromagnetic film by using a magnonic resonance-element formed by depositing a ferromagnetic stripe on top of the film. We found a magnonic counterpart to the Wood's anomaly as well as GH effect shift measurable with state-of-the-art experimental techniques. Our results have several important implications for magnonics and its application. First, our system is a platform for studying controlled reflection and scattering of SWs. Under certain conditions, the incident SW beam can excite an LM in the resonance element, which emits a part of its energy in the form of new SW beams. Note that, the resonance criterion is fulfilled in the system not for the specific but for a quite broad range of $M_\mathrm{S}$, covering a standard values of Py, indicating the feasibility of experimental realisation of such an interferometer. Moreover, our system allows for an easy change of the GH shift magnitude by several wavelengths, up to $450$~nm for the primary beam and $1600$~nm for the secondary beam. The exploited resonance is between being strictly confined to the stripe mode and a SW continuum band in the film. Thus, the same effects are expected for other beams' angles of incidence and other confined modes at different frequencies. Although the analysis in the main part of the paper was done for the changes in the value of stripe's $M_\mathrm{S}$,  we provide, in the supplementary material, the results with modulation of the frequency that revealed similar effects, which makes experimental realisation easier. Moreover, for the near-resonance scenario, the amplitude of the secondary beam can exceed the amplitude of the primary reflected beam even to a greater extent. Finally, the proposed geometry allows transfering energy of the SW beam propagating in the film to the stripe, i.e., it allows for high-efficiency transfer of SWs from 2D platforms into 1D waveguides, forming a transdimensional magnonic router in a similar manner to the one that was proposed for plasmons\cite{dong2020chip}. 
    This is crucial for designing magnonic circuits and exploiting third dimension for signal processing.

\begin{acknowledgments} 
    The research leading to these results has received funding from the Polish National Science Centre projects No. 2019/35/D/ST3/03729 and 2022/45/N/ST3/01844. 
    The numerical simulations were performed at the Poznan Supercomputing and Networking Center (Grant No. 398).
\end{acknowledgments}

\section{Supplementary}

 \subsubsection{Numerical methods}
    
        To perform numerical simulations, we employ the open-source environment Mumax3 \cite{vansteenkiste2014design}. This environment solves Landau-Lifshitz-Gilbert equation using the finite-difference method in the time domain. The simulated system has dimensions 12.7~\textmu m, 10.2~\textmu m and 15~nm (along the $x$, $y$, $z$ axis, respectively). We discretize the simulated domain with a regular mesh of unit cell $5\times 5 \times 5$~nm$^3$ (along the $x$, $y$, $z$ axes). In order to simulate an infinitely long system in $y$-axis and half-infinite system along the $x$-axis, we impose at all edges of the system except the one where the stripe is located, absorbing regions where the damping constant $\alpha$ increases quadratically to the value $\alpha_\mathrm{edge}=0.5$ at length of $L=625$~nm.
        
        We perform three types of simulations:
        \begin{itemize}
            \item calculations of the dispersion relation of the system for different values of stripe's $\it{M_\mathrm{S}}$,
            \item calculations of the steady-state for oblique incidence of continuously emitted spin-wave (SW) beam,
            \item reflection of a wave-packet with step-by-step observation of SWs reflection from the bilayer interface.
        \end{itemize}
        
        \subsubsection{Dispersion relation computations}
        
        To accelerate dispersion relation computation for SWs propagating along the stripe, 
        we perform simulations for a narrower system of width 1270~nm along the $x$-axis since, as we verified, it provides exactly the same results as simulations for the system of widths 12.7~\textmu m.  
        In this type of simulations, we place the SW source in the stripe parallel to the $x$-axis in the central part of the stripe. 
        To excite SWs for all wavevectors up to the cut-off wavevector $k_\mathrm{cut}=150$~rad/\textmu m and frequencies up to the cut-off frequency $f_\mathrm{cut}=20$~ GHz, we use the following spatial and temporal distribution of the microwave field being linearly polarized along the $z$-axis
        \begin{equation}
            \begin{split}
                h_z(t;x,y)=& h_0 \mathrm{sinc}(k_\mathrm{cut} y) \mathrm{sinc}(2\pi f_\mathrm{cut} (t-8/f_\mathrm{cut}) \\
                & \times \sum_{n=0}^N \big[ \mathrm{cos}(2\pi n\ x/w)+\mathrm{sin}(2\pi n x/w) \big], 
            \end{split}
            \label{eq:sinc_field}
        \end{equation}
        where the summation of $n$ is used to increase the efficiency of the higher order modes excitation (we assume $N=5$). We use the time sampling $t_{\mathrm{sampl}}~=~(2.2f_\mathrm{cut})^{-1}$ and save first $1000$ snapshots of the system's response to the microwave excitation. To obtain the dispersion relation $D(f,k_\mathrm{y})$ we employ following formula
        \begin{equation}
            D(f,k_{y}) = \langle |F_{t,y}\{m_{x}(t,y,x)\}| \rangle_{x \in \langle 0, \mathrm{w} \rangle},
            \label{eq:dispersion}
        \end{equation}
        where $F_{t,y}$ is the two-dimensional ($t,y$) fast Fourier transform (FFT), $m_\mathrm{x}(t,y,x)$ is the magnetic response taken only from the stripe. 
        The absolute value of the outcome of FFT ($|F_{t,y}\{m_{x}(t,y,x)\}|$) is a space-averaged along resonator's width $x \in \langle 0, \it{w} \rangle$ and represents $D(f,k_\mathrm{y})$.
    
        
        The results of the dispersion relation calculations for different values of stripe $M_\mathrm{S}$ are compiled into a short video that can be found in supplementary materials, Movie S3. 
        With an increase in the value of $M_\mathrm{S}$, the positions and shape of the dispersion relation bands change. 
        For easier analysis, the dashed lines indicate the parameters of the SW beam excited in the simulations. We show in the video that only in a specific range of $M_\mathrm{S}$ values, the bands cross the lines that represent the parameters of the incident SW beam. We interpret this range of $M_\mathrm{S}$ as a region of efficient SW excitation in the magnetic stripe.

        \subsubsection{Steady-state simulations}
        
        In order to excite SW beam, we use microwave magnetic field located at the left upper quarter of the layer.
        The spatial distribution of the dynamic magnetic field is in the rotated coordinate system $(x^\prime,y^\prime)$ by $45^{\circ}$ with respect to the $y$-axis. The spatio-temporal function of the dynamic magnetic field is given by a formula
        \begin{equation}
            \begin{split}
                B_{\mathrm{ext}, x}(t,x',y') = A(1-e^{-0.2 \pi f_0 t})R(x')G(y') \\
                \times [  \mathrm{sin}(k_0 x')\mathrm{sin}(2 \pi f_0 t) + \mathrm{cos}(k_0 x')\mathrm{cos}(2 \pi f_0 t) ],
            \end{split}
            \label{eq:sw}
        \end{equation}
        where $A=0.1B_0$ is the amplitude of the dynamic field ($B_0$ is the external magnetic field set along the system's $y$-axis of magnitude $B_0=0.01$~T), $R(x')=\Theta(-x'+\frac{w_a}{2})\Theta(x'+\frac{w_a}{2})$ is a rectangle function, which describes antenna's shape along its $x^\prime$ coordinate ($\Theta$ is Heaviside step function, antenna's width $w_a=30$~nm), $G(y')=\mathrm{exp}(-\frac{y'^2}{4\sigma_y^{2}})$ is a Gaussian function defining antenna's shape along the $y'$-axis ($\sigma_y=330$~nm), $k_0=60.96$~$\frac{\mathrm{rad}}{\mathrm{\mu m}}$ is the wavevector and $f_0=17.4$~GHz is the frequency of the excited SWs. Eq. (\ref{eq:sw}) enables unidirectional emission of SWs\cite{whitehead2019graded}. We use the antenna to constantly emit the SW beam for $41$~ns, after this time the system reaches the steady-state. Subsequently, we store time and space dependence of magnetization distribution for one period of SWs excitation in form of $25$ snapshots of magnetization distribution in the system with a sampling interval $1/(25 f_0)$. 
        
        The stored magnetization dependence on time in the steady-state can be converted into complex SW amplitude distribution at frequency $f_0$. It simplifies the analysis of the SW amplitude and phase.
        To make such a conversion, we calculate pointwise  FFT over time and select results only for $f_0$. 
        Furthermore, FFT analysis reveals that we obtain in the spectrum only one peak at $f_0$ what confirms linear response of the system. 

        \subsubsection{Simulations of the reflection of wave-packet}

        To simulate the wave-packet reflection, we use the same spatial distribution of the dynamic magnetic field as in Eq. (\ref{eq:sw}).  
        However, the time dependence of the formula is multiplied by the Gaussian envelope described by the expression $\mathrm{exp}(-(\frac{t}{2\sigma_f})^2)$, with $\sigma_f= 0.05 f_0$. It provides the packet with full width at half maximum (FWHM) in the time domain of $0.5$~ns. As the result of the simulations, we save $250$ snapshots of the propagating wave-packet with a time step of $0.057$~ns. The results of simulations with stripe's $M_\mathrm{S}=\left\{ 350, 550\right\}$~kA/m are compiled into short movies that can be found in supplementary materials, Movies S4 and S5. In the movie for the $M_\mathrm{S}=350$~kA/m stripe, the packet is reflected from the interface without any substantial excitation of the SWs in the stripe. However, in the movie with the $M_\mathrm{S}=550$~kA/m stripe, the excitation of the SWs in the stripe is evident. The mode formed in the stripe propagates along the stripe, and the re-emission of additional SWs to the layer is visible in the magnification.

        \subsubsection{Influence of the beam width on excitation of modes in the stripe}
        
        \begin{figure*}
            \centering
            \includegraphics[width=0.7\textwidth]{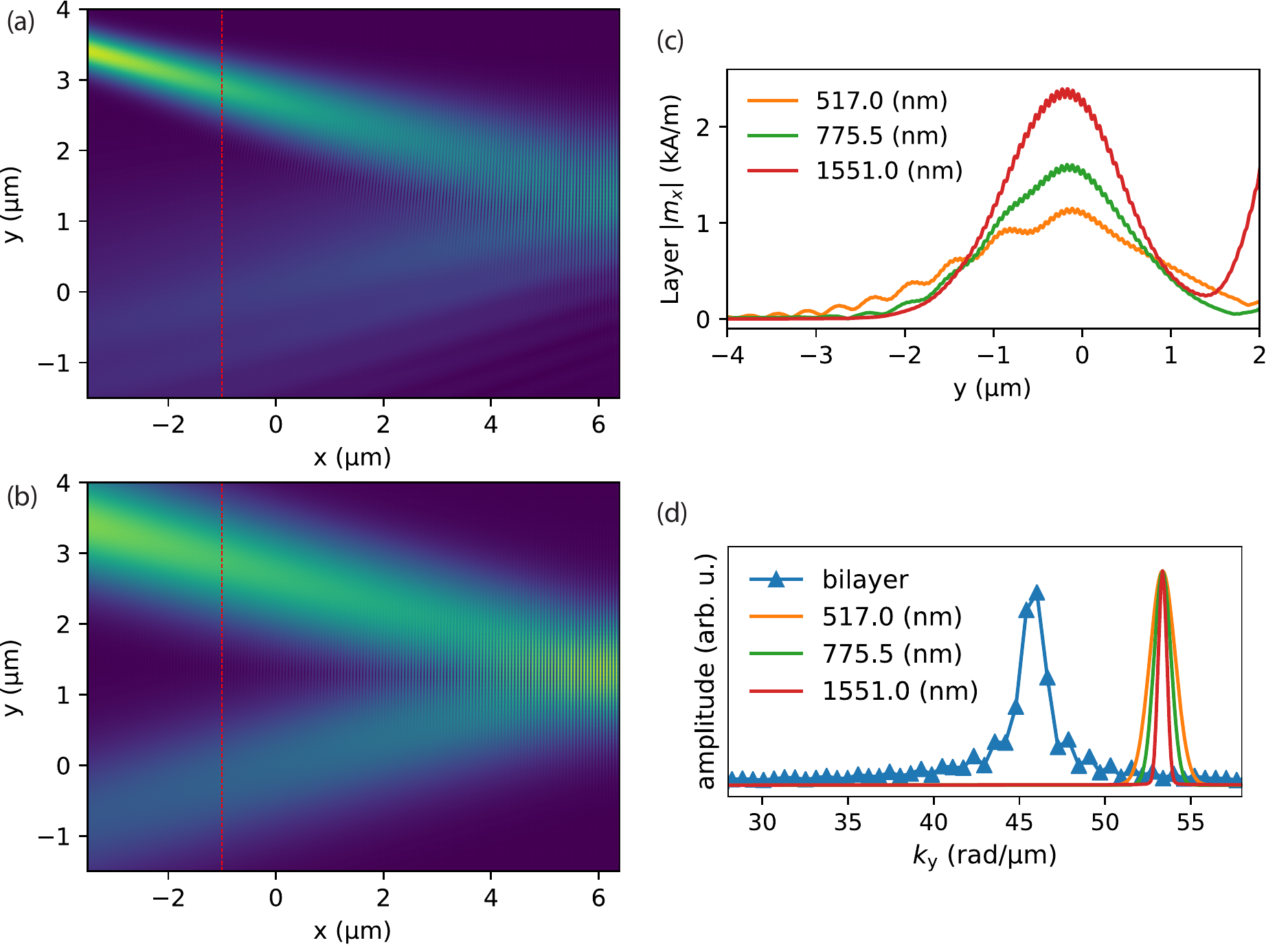}
            \caption{Beam width sweep. (a) SW intensity distribution in the layer with SW beam with FWHM $517$~nm at the antenna. (b) SW intensity distribution in the layer with SW beam with FWHM $1551$~nm at the antenna. (c) Comparision between SW intensity cutlines (marked with red dashed lines in (a,b)) for SW beam with FWHM $517, 775.5$ and $1551$~nm (orange, green and red lines respectively). The increase of the amplitude for the red line at $y>1.5$ \textmu m is caused by the widened incident beam. (d) Overlap of the beams width in the inverse space with the system's dispersion relation at $f=17.4$~GHz (blue line).}
            \label{fig:beam_width}
        \end{figure*} 
        
        We check the influence of the beam width on excitation of the modes in stripe for $M_\mathrm{S}=460$~kA/m as for this value of $M_\mathrm{S}$ for a beam with FWHM $=775$~nm, used in the main simulations, we observe the beginning of stripe's mode excitation. We perform series of simulations with beam of varying FWHM. In Fig. \ref{fig:beam_width} we show the results of simulations for beams with $\mathrm{FWHM}=\left\{517, 775.5, 1551\right\}$~nm. In Figs. \ref{fig:beam_width}(a,b) we present SW intensity distributions in the layer for two beams with $\mathrm{FWHM}=517$~nm (a) and $\mathrm{FWHM}=1551$~nm (b). In Fig. \ref{fig:beam_width}(a) with the narrow beam several reflected beams are evident. Their number is bigger than we presented in the main body of the paper. Conversely for a wide beam, as in Fig. \ref{fig:beam_width}(b), no reflected beam stratification is visible. For easier analysis in Fig. \ref{fig:beam_width}(c) we provide a plot with cutlines through SW intensity distributions for different SW beams, the cutlines are marked with red dashed lines in Figs. \ref{fig:beam_width} (a,b). The result of the narrowest beam is presented with the orange line. In this case the amplitude of the primary beam is the smallest but a phalanx of additional reflected beams are well visible. The results for the widest beam is shown with red line, here the reflected beam has regular Gaussian envelope and no additional reflected beams are visible. The green line presents the results for the beam with $\mathrm{FWHM}=775.5$~nm, used in the main simulations. This case is intermediary between previously presented narrow and wide beams. The secondary reflected beams are present although there are not as distinctive as in the case of the narrowest beam.
        
        We propose following explanation to the fact that narrow SW beam is able to excite the resonator's mode more efficiently than a wide beam. We bind the efficiency of mode excitation with an overlap between the beam's dispersion relation and the bilayer's dispersion relation. In Fig. \ref{fig:beam_width}(d) we show the cutlines through bilayer's dispersion relation at frequency $f=17.4$~GHz and dispersion relations of the beams. The dispersion relations of the beams are presented as Gaussian curves, which centres are calculated from the Kalinikos-Slavin formula \cite{kalinikos1986theory}, and their widths are obtained by calculating beam's widths in the reciprocal space. It is evident that the narrowest beam has the biggest width in the reciprocal space and because of that has the biggest overlap with bilayer's dispersion relation. Such a situation leads to more efficient coupling between the beam and the bilayer than in any other case presented in this analysis. Thus, the narrower beams have possibility to excite resonator's mode more efficiently in our system. The analysis presented here is more qualitative rather than quantitative as the beams widen during their propagation and during reflections have bigger FWHM than at the antenna. Thus the overlap of dispersion relation at the reflection is even smaller than presented in Fig. \ref{fig:beam_width}(d). However, the ratio between the overlap and the width of particular beam is the same as presented, so our explanation is justified. 
    
        \subsubsection{Results of comparision with Tamir's model}
        
        In this section we compare the results of our simulations with an analytical model proposed by Tamir and Bertoni\cite{tamir1971lateral} in more detail. Tamir and Bertoni showed that an incident beam of light is able to excite a leaky mode (LM) at the edge of the system. The excited edge mode propagates along the edge and emits waves back to the system. As our findings are a close analog to Tamir's model but in the realm of magnonics, we try to apply Tamir's mathematical description to our simulation's results. Tamir and Bertoni proposed the reflectance coefficient in the form of $\rho(k_y)=e^{i\Delta}(k_y-k_\mathrm{p}^{\ast})/(k_y-k_\mathrm{p})$, where $k_\mathrm{p}=\kappa + i \nu$ is a complex wave vector of the LM. They solved the system analytically under assumptions of well collimated beam incident and an angle of perfect coupling between the beam and edge mode. Additionally they also assumed that only the first pole in reflectance coefficient $\rho(k_y)$ provides a substantial input to the calculations. Their formula of the reflected light amplitude has two components which describe the primary $E_{0}$ and the secondary beam $E_{1}$
        \begin{equation} 
            \begin{split}
                E_{\mathrm{refl}} = E_{0} + E_{1}, \\  
                E_{0} = A e^{-((y-y_0)/w_b)^2}, \\
                E_{1} = -E_{0}(2 - \pi^{\frac{1}{2}} \nu w_b e^{(\gamma')^{2}} \mathrm{erfc}(\gamma')),
            \end{split}
            \label{eq:tamir_model}
        \end{equation} 
        where $A$ is the amplitude of the primary beam, $y_0$ is the centre of primary beam, $w_b$ is the width of beam at $\frac{1}{e}$ of its amplitude,  $\nu$ is the imaginary part of the LM wave-vector, $\gamma'$ is a new coordinate system defined for the secondary beam as $\gamma' = \frac{\nu w_b}{2} - \frac{y-y_0}{w_b}$ and $\mathrm{erfc}$ is the Gauss error function. 
        
        \begin{figure}
            \centering
            \includegraphics[width=0.4\textwidth]{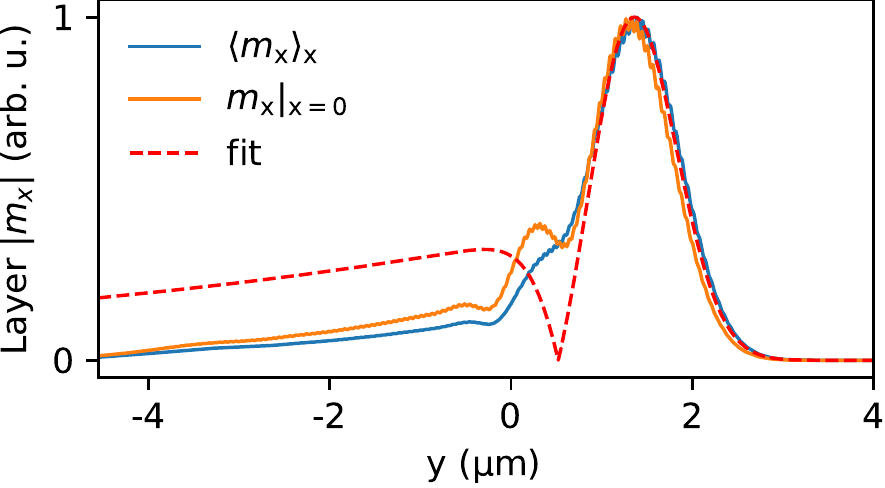}
            \caption{Comparision between simulation results (blue line - mean values of SW intensity in the layer under the stripe, orange line SW intensity cutline in the layer under the left edge of the stripe) and Tamir's model numerical fit (red dashed line) to the SW intensity mean value.}
            \label{fig:tamir}
        \end{figure}
        
       In Fig. (\ref{fig:tamir}) we present a numerical fit of Eq. (\ref{eq:tamir_model}) to the simulation data. The blue line in Fig. (\ref{fig:tamir}) shows the mean value of SW intensity averaged in the volume of the layer directly under the stripe. The dashed red line is the numerical fit to this data. It is evident that the analytical model provided by Tamir and Bertoni agrees only qualitatively with the results of our simulation. Namely, Tamir and Bertoni model describes properly the primary beam in our simulations but fails to precisely fit to the secondary beam. In this case analytical model only indicates separation between the primary and secondary reflected beams. However, it does not recreate the shape of secondary beam, it only shows a long tail of nonzero amplitude left to the primary beam. We see a several reasons why Tamir and Bertoni model does not work properly with our simulation's results. Firstly, Tamir and Bertoni model was developed for the light beam, which physics is governed by Helmholtz equation, while in our case we deal with SWs that are described by Landau-Lifshitz equation. Secondly, in Tamir and Bertoni model the edge of the system is infinitely narrow but in our simulations we regard the bilayer with finite width as an edge. We show the difference between the approach of wide and narrow edges in Fig. (\ref{fig:tamir}). Here, the blue line describes average SW intensity under the stripe and the orange line is a cutline through SW intensity in the layer under the left edge of the stripe. The results of the edge cutline have more distinctive peaks with bigger amplitudes nevetherless in our calculations we have to choose the average values of SW intensity to take into account contribution from whole bilayer width. At last Tamir and Bertoni model is based on several assumptions, such as choosing an optimal incident beam angle to couple with the edge mode, that are not met in our numerical simulations. We did not look for the ideal conditions for the SW beam incident in our simulations as their are impractical in designing experiments to confirm our numerical findings.

        \subsubsection{Influence of SW frequency on reflection}
        
\begin{figure}
            \centering
            \includegraphics[width=0.4\textwidth]{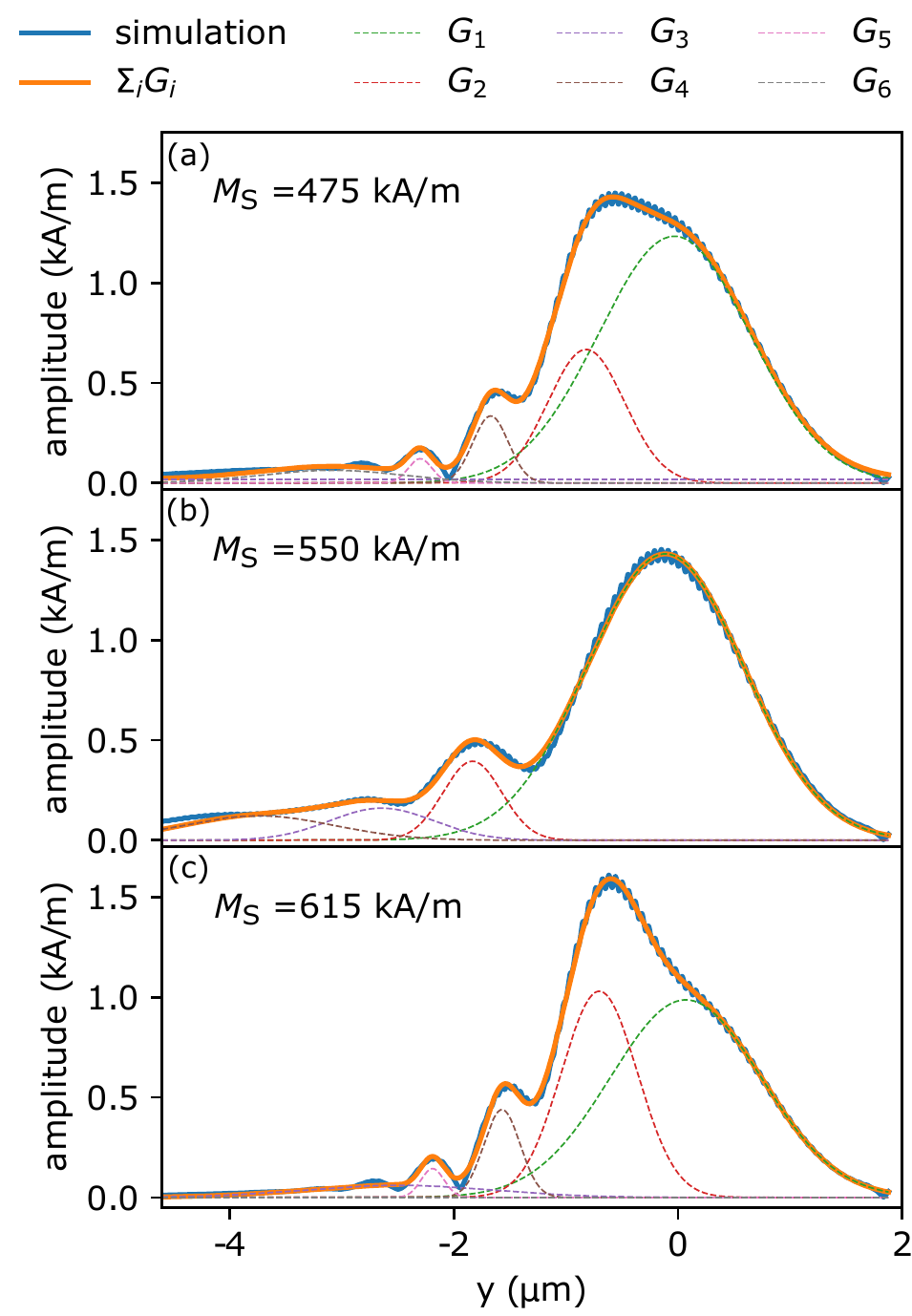}
            \caption{Calculating parameters of the reflected beams by fitting Gaussian curves in the far-field marked in Fig. \ref{fig:beam_width} with dashed-red line. (a) Fitting a sum of six Gaussian curves to the simulation results for resonator $M_\mathrm{S}=475$~kA/m. (b) Fitting a sum of four Gaussian curves to the simulation results for resonator $M_\mathrm{S}=550$~kA/m. (c) Fitting a sum of six Gaussian curves to the simulation results for resonator $M_\mathrm{S}=615$~kA/m.
            }
            \label{fig:fit}
        \end{figure}

        We obtain parameters of the reflected beams by fitting a sum of Gaussian functions to the cutline through SW intensity distribution in the far-field. The far-field is defined at $x=-7.5$ \textmu m and is indicated with a red dashed line in Figs. \ref{fig:beam_width}(a,b). We present our method of calculating reflected beams parameters in Fig. \ref{fig:fit}, where the blue solid line shows the simulation results, the dashed lines indicate component Gaussian functions and the orange solid line is the sum of all Gaussian curves in a given case. In the ranges of stripe's $M_\mathrm{S}$ when LM starts and ends to be excited the beams in the far-field strongly overlap as we show in Figs. \ref{fig:fit}(a,c) where $M_\mathrm{S}=475$~kA/m and $M_\mathrm{S}=615$~kA/m. In these cases we need to use a sum of six Gaussian functions to precisely fit our function to the simulation results. For the stripe's $M_\mathrm{S}$ values between these regions three distinctive beams and a range with plane waves are visible in the far-field as shown in Fig. \ref{fig:fit}(b) for $M_\mathrm{S}=550$~kA/m. Hence, we use as a fitting function the sum of four Gaussian curves only in this range of $M_\mathrm{S}$ (three to describe the beams and one to describe the plane waves). 

\begin{figure}
            \centering
            \includegraphics[width=0.4\textwidth]{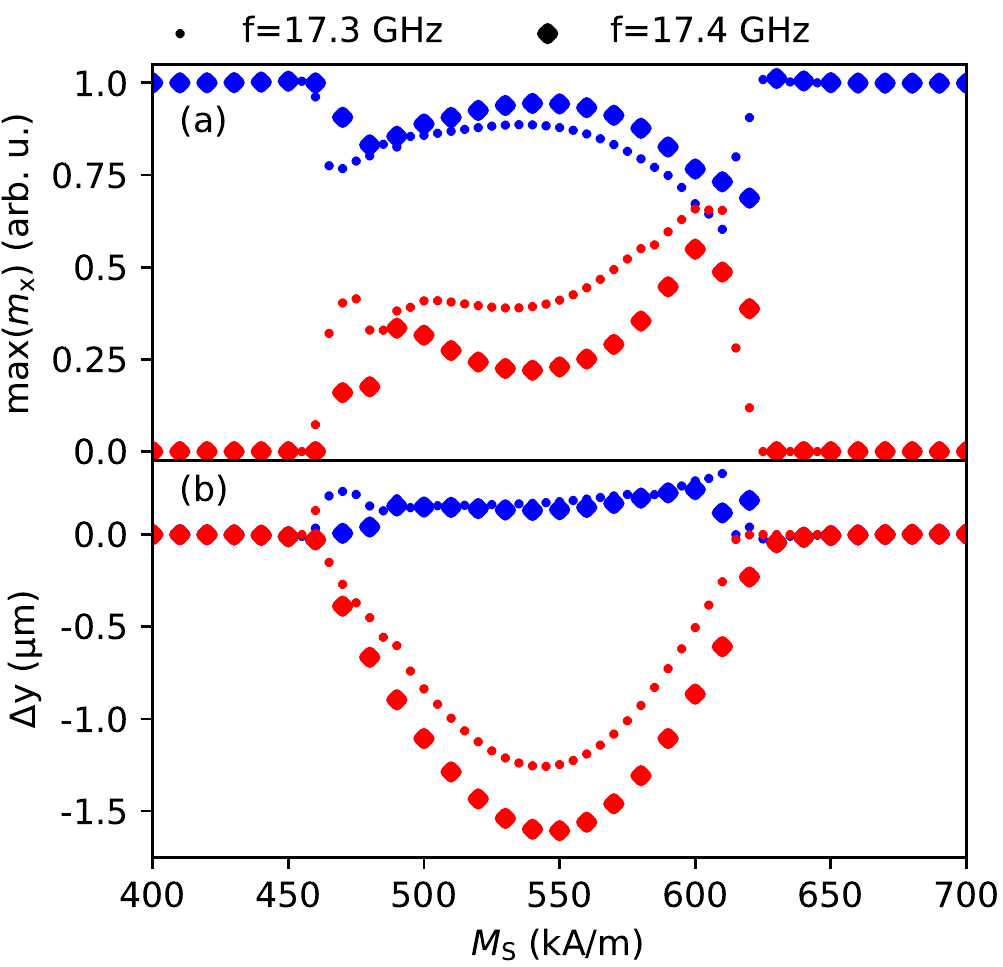}
            \caption{Parameters of the reflected beams in simulations with different SW frequencies. (a) Amplitudes of the primary (blue) and the secondary (red) beams calculated by fitting Gaussian functions to simulation data, cf. \ref{fig:fit}, dots represent results for frequency $f=17.3$~GHz, diamonds represent results for $f=17.4$~GHz. (b) Positions of the primary (blue) and the secondary (red) beams calculated by fitting Gaussian functions to simulation data, symbols as represent frequencies as in (a).  }
            \label{fig:amps}
        \end{figure}

        In Fig. \ref{fig:amps} we plot amplitudes and positions of the primary and secondary beams in the far-field as functions of stripe's $M_\mathrm{S}$. The blue colour represents parameters of the primary beams and the red colour depicts the secondary beam. Additionally in Fig. \ref{fig:amps} we also confront the results for simulations with different frequencies, namely the dots show results for $f=17.3$~GHz and the diamonds represent the results for $f=17.4$~GHz. The change in frequency affects overlap between dispersion relations of the SW beam and the bilayer thus changing the excitation properties of LM in the stripe. We chose only a small change in frequency to avoid bigger change of SW wavelength which would affect the wavelength-discretization ratio in the numerical simulations. The change of frequency in the system affects the amplitudes of reflected beams as presented in Fig. \ref{fig:amps}(a). The increase of frequency to $f=17.4$~GHz leads to increase of the primary beam's amplitude, compare blue dots and diamonds, and decrease of the secondary beam's amplitudes, compare red dots and diamonds. The same increase of frequency affect the spatial shift of the primary beam only slightly, as shown in Fig. \ref{fig:amps}(b) with blue dots and diamonds. However, the frequency increase causes substantial increase in spatial shift of the secondary beam, shown with red dots and diamonds. For $f=17.4$~GHz the maximal shift of the secondary beam is equal to $-1.6$ \textmu m and it is $0.35$ \textmu m bigger than spatial shift calculated for $f=17.3$~GHz and the same value of stripe's $M_\mathrm{S}$. 
        
        \subsubsection{Movie S1--steady-state with a sweep over resonator's $M_\mathrm{S}$ value}

        The movie.~S1 represents the colourmaps of the distribution of $|m_x|$ at frequency $f=17.4$ GHz as the dependence of the stripe's value of $M_\mathrm{S}$ similarly as displayed in Fig. 2(a,b). You can see that the distribution of $|m_x|$ is strongly affected by $M_\mathrm{S}$ and 3-6 reflected parallel laterally shifted beams can be seen depending on the $M_\mathrm{S}$ value.

        \subsubsection{Movie S2--steady-state with a sweep over frequency}
        
        The movie.~S2 depicts the colourmaps of the distribution of $|m_x|$ as the dependence on the value of frequency for resonator with $M_\mathrm{S}=550$ kA/m. We assume frequencies in the range from 17.0 to 18.0 GHz. Although, it is a narrow range, we can see that $m_x$ distribution changes significantly. While entering the resonance multiple reflected beams emerges and than again disappears. It is very similar result as for the sweep over resonator's $M_\mathrm{S}$. 
        
        \subsubsection{Movie S3--dispersion relation dependence on resonator's $M_\mathrm{S}$}
        
        The movie.~S3 depicts how the dispersion relation in the stripe depends on the value of stripe's $M_\mathrm{S}$. For the values of $M_\mathrm{S}$ from $420$~kA/m to $650$~kA/m we can see the crossing at $f=17.4$ GHz and $k_y=-53.5$ $\frac{\mathrm{rad}}{\mathrm{\mu m}}$ of the dispersion relation plotted for the stripe (colourmap in the background) and the dispersion plotted for SWs propagating in the layer with $\varphi=45^\circ$ (bold black line). 
        For this particular band crossing in the dispersion for SWs in the layer, at the considered  $M_{\mathrm{S}}$ range, the decrease in exchange energy is  compensated by an increase of the magnetostatic energy. Namely, the exchange energy is proportional to $M_{\mathrm{S}}^{-1}$, while the magnetostatic energy is proportional to $M_{\mathrm{S}}$. 
        It explains  the origin of the broad range of $M_{\mathrm{S}}$ showed in Fig. 4 in the main part of the manuscript where the resonance condition are fulfilled.

        \subsubsection{Movies S4-S5--reflection of wavepackets for different $M_\mathrm{S}$}
        
        The movie.~S4 shows the reflection of wavepacket from the resonant-stripe element in case of stripe $M_\mathrm{S}=350$~kA/m. In Movie.~S4 the antenna in simulation has width $w_a=200$~nm. The movie.~S5 shows SWs wavepacket reflection in case of stripe's $M_\mathrm{S}=550$~kA/m. 
        Comparing the results of wavepacket simulations for cases $M_\mathrm{S}=350$~kA/m and $M_\mathrm{S}=550$~kA/m it is evident that for latter the excitation of the SWs in the stripe is much more efficient. Without the constant SWs pumping by the SWs beam we can see propagation of a mode in the stripe as an obliquely bouncing between stripe's edges and reemition of SWs back to the layer clearly. Interestingly, in $M_\mathrm{S}=350$~kA/m stripe case we still are able to see excitation of a mode in the stripe and the SWs reemition, however with much smaller magnitude comparing to the case with $M_\mathrm{S}=550$~kA/m stripe. We explain this particular result by pointing that a SW wavepacket contains a range of frequencies described by a gaussian curve in our case centred at $f_0=17.4$~GHz with $\mathrm{FWHM}\approx2$~GHz. It means that the part of wavepacket spectrum still overlaps with the frequencies of resonant-stripe element's modes. This effect has small magnitude and is therefore virtually invisible in simulation results with continuous SW beam excitation.

\bibliography{literature}

\end{document}